\documentclass[prl, twocolumn, superscriptaddress,floatfix]{revtex4}%

\usepackage{amsfonts}
\usepackage{amsmath}
\usepackage{amssymb}
\usepackage{graphicx}
\usepackage{epstopdf}
\usepackage{textcomp}
\usepackage[normalem]{ulem}
\usepackage{color}
\usepackage{gensymb}

\begin{document}

\preprint{}

\title{Anisotropic superconductivity and magnetism in single-crystal RbEuFe$_4$As$_4$}

\author{M.~P.~Smylie}
\affiliation{Materials Science Division, Argonne National Laboratory, 9700 S. Cass Ave., Argonne, Illinois 60439}
\affiliation{Department of Physics, University of Notre Dame, Notre Dame, Indiana 46556}
\author{K. Willa}
\affiliation{Materials Science Division, Argonne National Laboratory, 9700 S. Cass Ave., Argonne, Illinois 60439}
\author{J.-K. Bao}
\affiliation{Materials Science Division, Argonne National Laboratory, 9700 S. Cass Ave., Argonne, Illinois 60439}
\author{K. Ryan}
\affiliation{Department of Physics, University of Illinois at Urbana-Champaign, Urbana, Illinois, 61801, USA}
\author{Z. Islam}
\affiliation{Advanced Photon Source, Argonne National Laboratory, 9700 S. Cass Ave, Lemont, Illinois, 60439, USA}
\author{H. Claus}
\affiliation{Materials Science Division, Argonne National Laboratory, 9700 S. Cass Ave., Argonne, Illinois 60439}
\author{Y. Simsek}
\affiliation{Materials Science Division, Argonne National Laboratory, 9700 S. Cass Ave., Argonne, Illinois 60439}
\author{Z. Diao}
\affiliation{Laboratory of Mathematics, Physics and Electrical Engineering, Halmstad University, SE-301 18 Halmstad, Sweden}
\affiliation{Department of Physics, Stockholm University, SE-106 91 Stockholm, Sweden}
\author{A. Rydh}
\affiliation{Department of Physics, Stockholm University, SE-106 91 Stockholm, Sweden}
\author{A. E. Koshelev}
\affiliation{Materials Science Division, Argonne National Laboratory, 9700 S. Cass Ave., Argonne, Illinois 60439}
\author{W.-K. Kwok}
\affiliation{Materials Science Division, Argonne National Laboratory, 9700 S. Cass Ave., Argonne, Illinois 60439}
\author{D. Y. Chung}
\affiliation{Materials Science Division, Argonne National Laboratory, 9700 S. Cass Ave., Argonne, Illinois 60439}
\author{M. G. Kanatzidis}
\affiliation{Materials Science Division, Argonne National Laboratory, 9700 S. Cass Ave., Argonne, Illinois 60439}
\affiliation{Department of Chemistry, Northwestern University, Evanston, Illinois, 60208, USA}
\author{U. Welp}
\affiliation{Materials Science Division, Argonne National Laboratory, 9700 S. Cass Ave., Argonne, Illinois 60439}

\begin{abstract}

We investigate the anisotropic superconducting and magnetic properties of single-crystal RbEuFe$_4$As$_4$ using magnetotransport and magnetization measurements.
We determine a magnetic ordering temperature of the Eu-moments of $T_m$ = 15 K and a superconducting transition temperature of $T_c$ = 36.8 K.
The superconducting phase diagram is characterized by high upper critical field slopes of -70 kG/K and -42 kG/K for in-plane and out-of-plane fields, respectively, and a surprisingly low superconducting anisotropy of $\Gamma$ = 1.7.
Ginzburg-Landau parameters of $\kappa_c \sim 67$ and $\kappa_{ab} \sim 108$ indicate extreme type-II behavior.
These superconducting properties are in line with those commonly seen in optimally doped Fe-based superconductors.
In contrast, Eu-magnetism is quasi-two dimensional as evidenced by highly anisotropic in-plane and out-of-plane exchange constants of 0.6 K and $<$ 0.04 K.
A consequence of the quasi-2D nature of the Eu-magnetism are strong magnetic fluctuation effects, a large suppression of the magnetic ordering temperature as compared to the Curie-Weiss temperature, and a cusp-like anomaly in the specific heat devoid of any singularity. Magnetization curves reveal a clear magnetic easy-plane anisotropy with in-plane and out-of-plane saturation fields of 2 kG and 4 kG.

\end{abstract}

\date{\today}

\maketitle

\begin{center}
\textbf{I. INTRODUCTION} 
\end{center}

\begin{figure}
	\includegraphics[width=0.8\columnwidth]{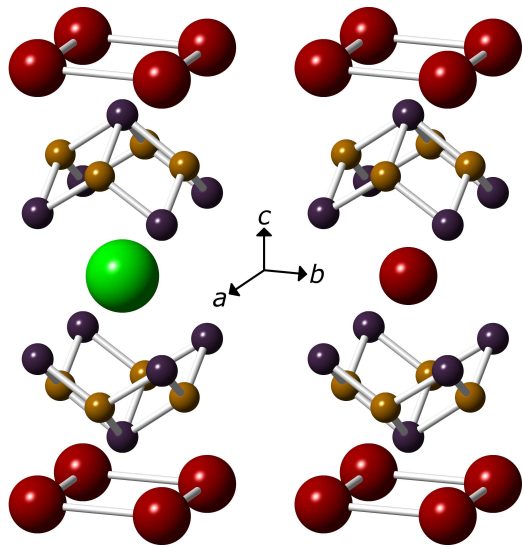}
	\caption{
		(L) Crystal structure (P4/mmm) of RbEuFe$_4$As$_4$, with 2D sheets of Rb (green) and Eu (red) separated by Fe$_2$As$_2$ blocks.
		(R) Crystal structure (I4/mmm) of the parent compound EuFe$_2$As$_2$.
	}
	\label{figStructure}
\end{figure}

Europium-containing Fe-based superconducting materials have emerged as model systems for the study of the interplay of magnetism and superconductivity \cite{ZapfDressel2017_review,CaoXu_JoPCS_SC-and-FM-in-pnictides_2012}.
They are the latest members of a family of superconductors in which superconductivity coexists with complete, magetically ordered sublattices of local rare-earth ($R$) moments such as $R$Rh$_4$B$_4$ \cite{MapleFischer1982}, $R$Mo$_8$S$_8$ \cite{FischerMaple1990,Canfield1998_PhysToday}, and the nickel borocarbides \cite{MullerNarozhnyi_RepProgPhys2001}.
It is believed that in these compounds the magnetic moments and the superconducting electrons reside in different, essentially isolated sublattices, enabling the existence of superconductivity despite the high concentration of localized magnetic moments \cite{MullerNarozhnyi_RepProgPhys2001,Eisaki1994_PhysRevB.50.647}.
Among these, the europium-containing Fe-based superconducting materials stand out since they display simultaneously high magnetic ordering temperatures (15-20 K) and superconducting transition temperatures in excess of 30 K, implying sizable magnetic exchange interactions in the presence of strong superconducting pairing.
Extensive work on EuFe$_2$As$_2$ (Eu-122) derived compounds has shown that the non-superconducting parent compound undergoes a spin density wave (SDW) transition of the Fe-magnetic moments near 195 K \cite{Jeevan2008_PhysRevB.78.052502} and near $T_m \sim 19$ K a transition of the Eu-moments into a type-A antiferromagnetic state in which ferromagnetically ordered Eu-sheets are coupled antiferromagnetically along the $c$-axis \cite{Xiao2009_PhysRevB.80.174424}.
A similar magnetic structure has been found in the low-temperature phases of the Ho, Dy, and Pr-borocarbides \cite{Lynn1997_PhysRevB.55.6584}.
Upon the application of pressure \cite{Miclea2009_PhysRevB.79.212509,Matsubayashi2011_PhysRevB.84.024502} or doping with, among others, P \cite{RenXu2009-PhysRevLett.102.137002,Jeevan2011_PhysRevB.83.054511}, K \cite{Jeevan2008_PhysRevB.78.092406,Maiwald2012_PhysRevB.85.024511,Anupam2011_JoPCM}, and Na \cite{QiYWMa2008_NJoP-(EuNa)Fe2As2,QiYWMa2012_NJoP-(EuNa)Fe2As2}, the SDW transition of Eu-122 is suppressed and superconductivity emerges at temperatures reaching up to $T_{c} \sim$ 30 K.
At the same time, the Eu-moments in the case of EuFe$_2$(As$_{1-x}$P$_x$)$_2$ rotate from the $ab$-plane close to the $c$-axis \cite{Nandi2014_PhysRevB.90.094407}; however, $T_c$ stays largely unaltered inside the superconducting dome.
This apparent decoupling of the magnetic Eu-sublattice from the superconducting electrons has been attributed to the multi-orbital nature of the Fe-based superconductors in which magnetic exchange interactions and superconductivity are mediated by different groups of electrons and to the high upper critical fields that can withstand internal exchange and dipolar fields \cite{CaoXu_JoPCS_SC-and-FM-in-pnictides_2012,Cao2011_JoPCM}.
In addition, due to the crystal structure of EuFe$_2$As$_2$-based materials (see Fig.~\ref{figStructure}) partial cancellation of exchange and dipolar fields may arise at the location of the Fe-atoms.

In this regard, the recent discovery of superconductivity in RbEuFe$_4$As$_4$ and CsEuFe$_4$As$_4$ \cite{KawashimaIwo2016_JPSJ-polycrystalline-RbEu1144,IyoYoshida2016_JACS-AeAFeAs4-discovery,LiuCao2016_PRB-polycrytalline-RbEu1144,LiuXuCao2016-SciBull} is significant since in these materials the asymmetric environment of the Fe$_2$As$_2$-layers (see Fig.~\ref{figStructure}) precludes any cancellation effects.
Nevertheless, $T_{c}$ reaches 37 K, among the highest values of all 122-type materials, and exceeds the values of the non-magnetic sister compounds CaKFe$_4$As$_4$ ($T_{c}$ = 35 K) \cite{MeierCanfield2016_PhysRevB.94.064501} and (La,Na)(Cs,Rb)Fe$_4$As$_4$ ($T_{c} \sim$ 25 K) \cite{Kawashima2018_JPCL-1144}.
This is in contrast to the behavior of nickel borocarbides for which the incorporation of magnetic rare earth ions leads to a clear suppression of $T_c$ as compared to a non-magnetic rare earth ion \cite{MullerNarozhnyi_RepProgPhys2001}.
RbEuFe$_4$As$_4$ and CsEuFe$_4$As$_4$ are intrinsically doped to 0.25 holes/Fe-atom such that in the stoichiometric material an electronic structure arises that closely corresponds to optimally doping in 122-materials.
Furthermore, a recent study \cite{LiuCao2017-PhysRevB.96.224510} revealed that upon Ni-substitution on the Fe-site $T_{c}$ is suppressed to zero and the SDW re-emerges, while at the same time $T_m$ is unchanged; similarly, Ca-substitution on the Eu-site \cite{Kawashima2018-LT28Proceedings} suppresses $T_{m}$ without changing $T_{c}$, demonstrating the almost complete  of the Eu-sublattice from superconductivity.

Here we present the first study of the anisotropic superconducting and magnetic properties of single-crystal RbEuFe$_4$As$_4$.
Using magnetotransport and magnetization measurements, we determine a magnetic ordering temperature of the Eu-moments of $T_m$ = 15 K and a superconducting transition temperature of $T_{c}$ = 36.8 K.
The superconducting phase diagram is characterized by high upper critical field slopes of $dH_{c2}^{ab}/dT = -70 $kOe/K, $dH_{c2}^{c}/dT = -42 $kOe/K,  and a surprisingly low superconducting anisotropy of $\Gamma$ = 1.7.
Ginzburg-Landau (GL) parameters of $\kappa_c \sim$ 67 and $\kappa_{ab} \sim$ 108 indicate extreme type-II behavior.
These superconducting properties are in line with those commonly seen in optimally doped Fe-based superconductors.
In contrast, Eu-magnetism is highly anisotropic quasi-two dimensional as evidenced by anisotropic in-plane and out-of-plane exchange constants of 0.6 and $<$ 0.04 K, respectively.
A consequence of the quasi-2D nature of the Eu-magnetism are strong magnetic fluctuation effects, a negative magnetoresistance in high fields and at temperatures well above $T_{c}$, a large suppression of the magnetic ordering temperature as compared to the Curie-Weiss temperature, and a cusp-like anomaly in the specific heat.
Magnetization curves reveal a clear magnetic easy-plane anisotropy with in-plane and out-of-plane saturation fields of 2 kOe and 4 kOe, respectively.

\begin{center}
\textbf{II. EXPERIMENTAL METHODS} 
\end{center}

High quality single crystals of RbEuFe$_4$As$_4$ were grown using RbAs flux \cite{Bao2018} yielding thin flat plates with sizes of up to 0.8mm x 0.8mm x 60 $\mu$m with the tetragonal $c$-axis (001) perpendicular to the plate and the tetragonal (110) and (1$\bar{1}$0) orientations parallel to the edges.
For magnetotransport measurements, thin bars were cut from plates and gold wires were then attached with silver epoxy onto bar-shaped samples in a standard 4-point configuration.
For $c$-axis current measurements, sets of two contacts were placed on the top and bottom faces of the single crystal, roughly equally spaced.
Magnetotransport measurements were performed in a 90-10-10 kG three-axis superconduting vector magnet avoiding the need for mechanically rotating or remounting the samples, and magnetization measurements were performed on both zero field-cooling (ZFC) and field-cooling (FC) in a Quantum Design MPMS-7 system with samples attached to a quartz rod or quartz fiber.
The specific heat of RbEuFe$_4$As$_4$ single crystals was measured using a membrane-based ac-nanocalorimeter \cite{TagliatiRydh_2012_RevSciInst-Nanocal,WillaRydh_2017_RevSciInst-Nanocal}.

X-ray diffraction (XRD) and specific heat measurements revealed single-phase material without EuFe$_2$As$_2$ inclusions \cite{Bao2018}.
At room temperature, RbEuFe$_4$As$_4$ has a simple-tetragonal crystal structure (P4/mmm space group) with one formula unit per unit cell and lattice constants of $a$ = 3.882 \AA~and $c$ = 13.273 \AA~(see Fig.~\ref{figStructure}).
The large difference in ionic sizes of the Eu and Rb ions induces their segregation into sheets.
The formal valence count reveals that RbEuFe$_4$As$_4$ is intrinsically doped to 0.25 holes/Fe-atom.
In contrast, the EuFe$_2$As$_2$ parent compound is at room temperature body-centered-tetragonal (I4/mmm space group) containing two formula units per tetragonal unit cell.

\begin{center}
\textbf{III. RESULTS AND DISCUSSION} 
\end{center}

\begin{center}
\textbf{A. Specific Heat}
\end{center}

\begin{figure}
	\includegraphics[width=1\columnwidth]{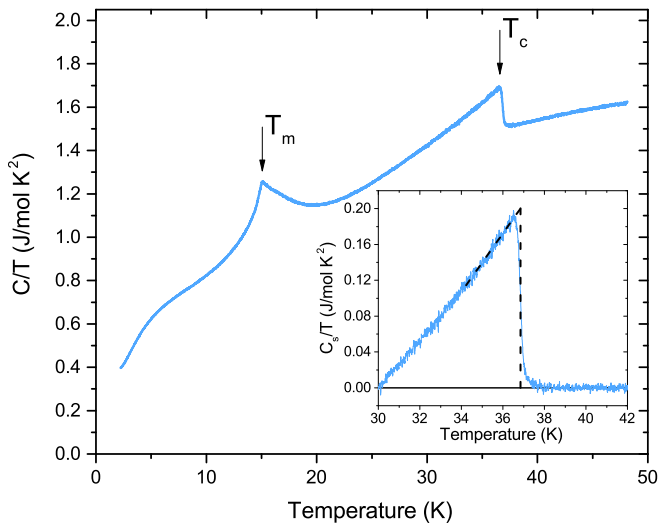}
	\caption{
		Temperature dependence of $C/T$ in zero-field specific heat of RbEuFe$_4$As$_4$.
		A clearly resolved cusp at $T_m$ = 15 K and a step at $T_{c}$ = 36.8 K mark the magnetic and superconducting transitions, respectively.
		The inset shows the superconducting specific heat near the transition on enlarged scales.
		The lines illustrate the entropy conserving construction.
	}
	\label{figCT}
\end{figure}

We evaluate the phase transitions occuring in single-crystal RbEuFe$_4$As$_4$ using zero-field specific heat measurements such as shown in Figure~\ref{figCT} \cite{Bao2018}.
A clear cusp in $C/T$ at $T_m \sim$ 15 K signals the magnetic transition whereas a step in $C/T$ at $T_{c} \sim$ 36.8 K is the signature of the superconducting transition.
Our samples do not display an additional feature in the specific heat near 5 K that has been reported on polycrystalline samples \cite{LiuCao2016_PRB-polycrytalline-RbEu1144,LiuXuCao2016-SciBull}, and was interpreted as signature of a transformation of a Fulde-Ferrell-Larkin-Ovchinnikov state into a spontaneous vortex state.
We observe a fairly large step size of $\Delta C/T_{c}$ = 0.21 J/mol K$^2$ at the superconducting transition as determined from the entropy conserving construction (inset of Fig.~\ref{figCT}).
In single-band weak-coupling BCS theory this step size correspods to a large coefficient of the normal state electronic specific heat of $\gamma_n = \Delta C/1.43T_{c}$ = 147 mJ/mol K$^2$.
Similar values have recently been reported for polycrystalline RbEuFe$_4$As$_4$ samples \cite{LiuCao2016_PRB-polycrytalline-RbEu1144} as well as for crystals of the non-magnetic sister-compound CaKFe$_4$As$_4$ \cite{MeierCanfield2016_PhysRevB.94.064501}.
In single-band weak-coupling BCS theory the normalized discontinuity of the slopes of the specific heat at $T_{c}$, $(T_{c}/\Delta C)\Delta(dC/dT)_{T_{c}}$, has a universal value of 2.64.
Strong-coupling and multi-band effects modify this value as seen for example in Pb for which a slope discontinuity of 4.6 has been reported \cite{Carbotte1990_RevModPhys.62.1027} and the two-band superconductor MgB$_2$ for which a value of 3.35 can be deduced \cite{Bouquet2001_PhysRevLett.87.047001}, respectively.
From the data in Fig.~\ref{figCT} we obtain a very large value of $(T_{c}/\Delta C)\Delta(dC/dT)_{T_{c}} \sim$ 6.9 which is similar to Ba$_{1-x}$K$_x$Fe$_2$As$_2$ \cite{Welp2009_PhysRevB.79.094505} and indicative of strong coupling effects.

The cusp-like feature at the magnetic transition does not display signatures commonly associated with a second order transition, $i.e.$, a step such as seen at the superconducting transition, or a singularity.
This observation is in agreement with previous reports on polycrystalline RbEuFe$_4$As$_4$ \cite{LiuCao2016_PRB-polycrytalline-RbEu1144} and CsEuFe$_4$As$_4$ \cite{LiuXuCao2016-SciBull} samples, where it has been attributed to a 3rd order phase transition.
It is however in contrast to EuFe$_2$As$_2$, which shows typical singular behavior in the specific heat at the magnetic transition \cite{Jeevan2008_PhysRevB.78.052502,RenXu2009-PhysRevLett.102.137002,RenXu2008-PhysRevB.78.052501,Paramanik2014_SST-Eu(FeIr)2As2,Oleaga2014_JoAaC-EuFe2As2}.

As shown in Fig.~\ref{figStructure}, an important difference between RbEuFe$_4$As$_4$ and EuFe$_2$As$_2$ is that in RbEuFe$_4$As$_4$ the distance between Eu-layers is twice as large as in EuFe$_2$As$_2$ suggesting that reduced dimensionality and strong fluctuation effects lead to the marked difference in the specific heat signatures.
In fact, due to the highly anisotropic exchange constants and the easy-plane magnetic anisotropy described in more detail below, the magnetism of Eu may be quasi-2D in RbEuFe$_4$As$_4$ exhibiting 2D-XY criticality and Berezinskii-Kosterlitz-Thouless behavior \cite{Tobochnik1979_PhysRevB.20.3761,Gupta1992_PhysRevB.45.2883} while in EuFe$_2$As$_2$ it is more 3D-like, and more accurately described by a 3D-XY model accompanied by a singular specific heat as seen in experiment.
Nevertheless, the measured cusp in the specific heat (Fig.~\ref{figCT}) is too sharp as compared to the predictions of the 2D-XY model \cite{Gupta1992_PhysRevB.45.2883}.
However, Monte Carlo simulations of the 2D to 3D crossover in the XY model clearly reveal the re-emergence of the singularity in the specific heat with increasing 3D-coupling \cite{Janke1990_PhysRevB.42.10673} indicating quasi-2D behavior in the data of Fig.~\ref{figCT}.
Similarly, cusp-like specific heat transitions arise in quasi-2D anisotropic Heisenberg models that depend on the coupling strength in the third direction \cite{Sengupta2003_PhysRevB.68.094423}.
A detailed examination of these phenomena is currently underway.

On decreasing temperature, the $C/T$ data display a pronounced downward curvature.
This feature, not seen on samples whose specific heat is dominated by the electronic and phonon contributions, has been reported for various high-spin systems \cite{Johnston2011_PhysRevB.84.094445,Bouvier1991_PhysRevB.43.13137}.
It does not represent a phase transition, but qualitatively, it arises from the crossover from the quantum regime at low temperatures for which $C$ approaches zero at zero temperature to the classical regime in which $C(T = 0)$ would be finite. 
This crossover is particularly sharp in high-$S$ systems since these follow classical behavior over most of the temperature range, and it is absent in $S$=1/2 systems as these are purely quantum mechanical.  

\begin{center}
\textbf{B. Magnetic Properties}
\end{center}

We determine the magnetic state of RbEuFe$_4$As$_4$ using measurements of the field cooled (FC) and zero-field cooled (ZFC) temperature dependence as well as the field dependence of the magnetization in fields applied along the $ab$-planes and the $c$-axis.
In contrast to EuFe$_2$As$_2$, the magnetic transition of the Eu-ions occurs deep in the superconducting state.
Therefore, magnetization data at low temperatures, especially ZFC data and data for which $H // c$, contain contributions from superconducting vortices as well as from Eu-moments.

\begin{figure}
	\includegraphics[width=1\columnwidth]{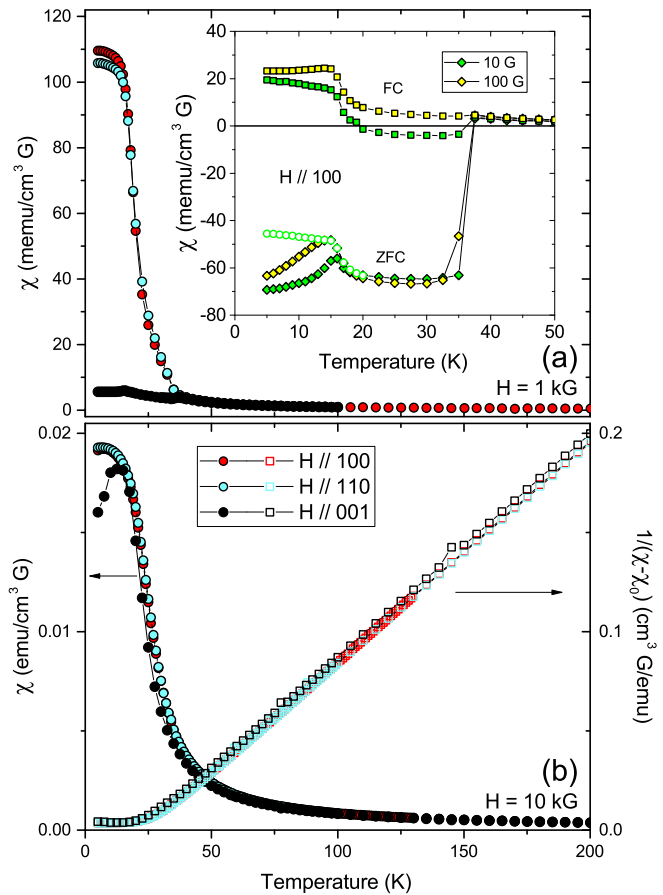}
	\caption{
		Inset of panel (a): Temperature dependence of the susceptibility measured on FC (solid squares) and ZFC (solid diamonds) in 10 G (green symbols) and 100 G (yellow symbols) applied along the (100) direction.
		The open green circles are FC data in 10 G after warming to 20 K.
		Panels (a) and (b): Temperature dependence of the susceptibility after field cooling in a field of 1 kG and 10 kG along the (100), (110), and (001) directions.
		At temperatures above 50 K, the data are well described by the Curie-Weiss expression $\chi(T) = \chi_0 + C/(T-\Theta_C)$.
	}
	\label{figXT}
\end{figure}

The inset of Figure~\ref{figXT}(a) shows the temperature dependence of the magnetic susceptibility, $\chi = M/H$, measured in FC and ZFC conditions in several fields applied parallel to the in-plane (100)-direction.
The large diamagnetic signal observed near 37 K in the ZFC data marks the superconducting transition.
The magnetic transition is seen as a clear cusp near $T_m$ = 15 K most notably in the ZFC data whereas on field cooling, the susceptibility attains an almost temperature independent value at the magnetic transition.
In the case of EuFe$_2$As$_2$, similar magnetization behavior has been shown to arise from a transition into a type-A antiferromagnetic state.
We note, however, that magnetization data such as shown in Fig.~\ref{figXT} do not allow to determine the actual magnetic structure of RbEuFe$_4$As$_4$.
For instance, EuCo$_2$P$_2$, which has the same crystal structure as Eu-122, displays magnetic behavior similar to that in Fig.~\ref{figXT}, although a helical antiferromagnetic structure has been proposed for this material \cite{Sangeetha2016-PhysRevB.94.014422}.
Also included in the inset of Fig.~\ref{figXT}(a) are data (green open circles) obtained following FC in 10 G after the sample was warmed on a ZFC run in 10 G up to 20 K showing that it is not required to pass through the superconducting transition in order to induce the ferromagnetic-like state.
We note that, in general, this FM-like state is induced on field-cooling in relatively low fields indicating a fragility of the pristine AFM order deep within the SC phase.
The main panel in Fig.~\ref{figXT}(a) displays the temperature dependence of the susceptibility measured after field cooling in a field of 1 kG applied along the three crystal axes.
Under FC conditions for which the effects due to vortex pinning are small, we observe a large anisotropy in the low temperature susceptibility with $\chi_{ab} >> \chi_c$ revealing a pronounced easy-plane anisotropy of the Eu-moments, similar to EuFe$_2$As$_2$.
The data also show that a possible in-plane magnetic anisotropy is comparatively very weak. 

In higher fields [Fig.~\ref{figXT}(b)], the difference between $\chi_{ab}$ and $\chi_c$ diminishes indicating that magnetic saturation is approached.
Below the ordering temperature, $\chi_c$ slightly decreases with decreasing temperature because the growing magnetic anisotropy pulls the Eu-moments towards the planes.
The data above 50 K are well described by a Curie-Weiss law $\chi(T) = \chi_0 + C/(T-\Theta_C)$ yielding for the (100), (110), and (001) directions values of $\Theta_C$ of 24.18 K, 23.81 K and 22.32 K, respectively, and values for $C$ of 7.476, 7.524 and 7.404 emu K/mol G, respectively.
With $\mu_{eff} = 2.827 * C^{1/2}$, we find an effective moment of $\sim7.75~\mu_B$ per Eu-ion.
This value is close to the expected Eu$^{2+}$ effective moment of $\mu_{eff} = g\mu_B \sqrt{S(S+1)} = 7.94~\mu_B/$Eu (with $g$ = 2 and $S$ = 7/2), indicating that essentially all Eu-ions are in the 2$^+$ state.
The positive value of the Curie-Weiss temperature signals predominantly ferromagnetic interactions between the Eu-moments consistent with a type-A antiferromagnetic structure.
We observe a sizable reduction of the magnetic ordering temperature of RbEuFe$_4$As$_4$ ($T_m$ = 15 K, $\Theta_C$ = 23 K) as compared to Eu-122 for which a Curie-Weiss temperature of $\Theta_C \sim$ 21 K and a magnetic ordering temperature $T_m \sim$ 19 K have been determined \cite{Jiang2009_NJoP}.
We attribute this difference to the reduced dimensionality and strong magnetic fluctuations in RbEuFe$_4$As$_4$, while in Eu-122, which has the same layered spin arrangement albeit with half the distance of that in RbEuFe$_4$As$_4$, magnetic fluctuations have a relatively reduced effect consistent with the more conventional form of the specific heat anomaly as discussed above.

The values for $\chi_0$ are $3.4 \times 10^{-3}$ emu/mol G for the in-plane orientations and $3.1 \times 10^{-3}$ emu/mol G for the $c$-axis representing anisotropic contributions from temperature-independent Pauli paramagnetism or van Vleck magnetism. 

\begin{figure}
	\includegraphics[width=1\columnwidth]{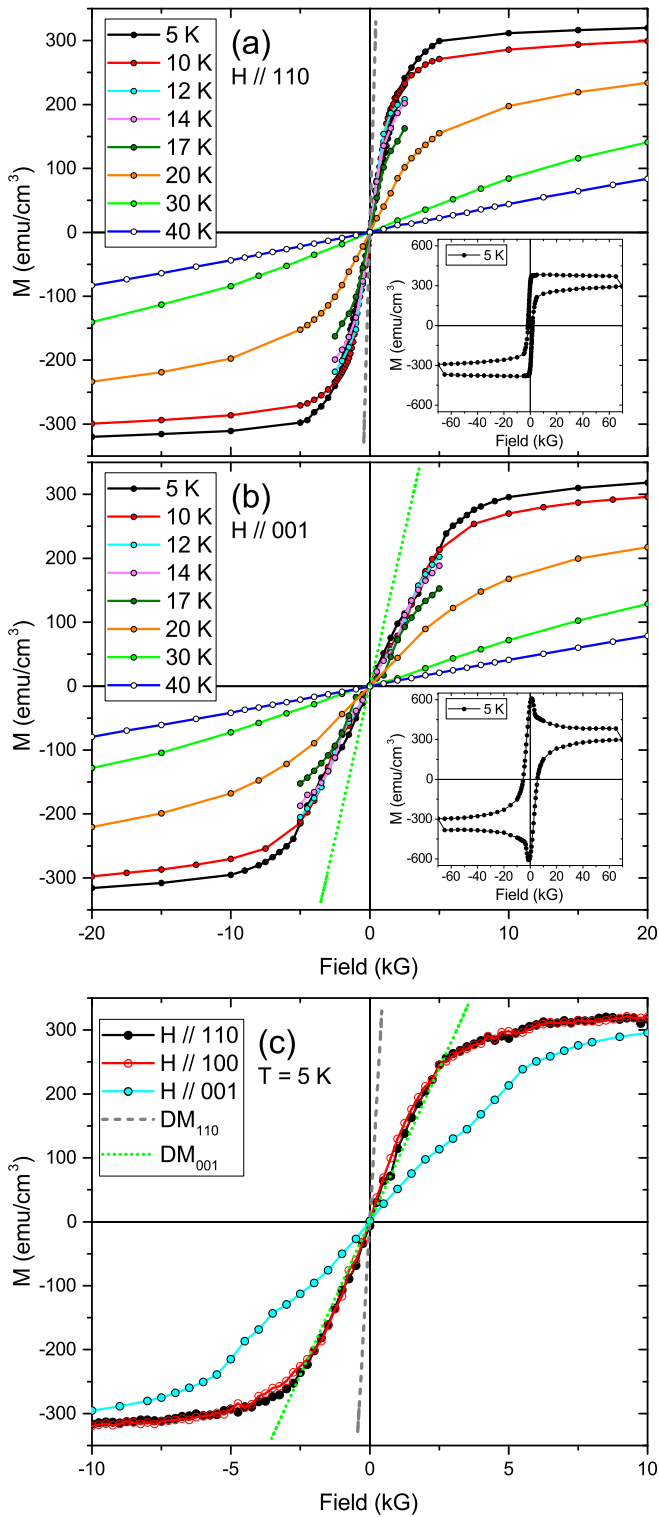}
	\caption{
		Magnetization of the Eu-sublattice vs applied field for (a) $H$ // (110) and (b) $H$ // (001) at various temperatures.
		The insets in (a) and (b) show the as-measured magnetization hysteresis loops.
		(c) shows a comparison of the magnetization at 4.5 K measured along the three crystal axes.
		The dashed lines represent the demagnetization fields due to the plate-like sample geometry.  
	}
	\label{figHysteresis}
\end{figure}

The insets of Figs.~\ref{figHysteresis}(a) and~\ref{figHysteresis}(b) show magnetization hysteresis loops measured at 4.5 K in $H$ // (001) and $H$ // (110).
The superposition of a ferromagnetic-like signal and a hysteretic superconducting signal is clearly seen, especially for $H$ // (001).
This is expected due to the large sample cross-section and high critical current density for this field orientation.
Assuming that the superconducting hysteresis is symmetric around the equilibrium magnetization curve and that effects due to the hysteresis of the Eu-magnetism are small (as is indicated by results on EuFe$_2$As$_2$ \cite{Jiang2009_NJoP}) we extract the magnetization curve of the Eu-sublattice as ($M_+ + M_-)/2$ where $M_+~(M_-)$ is the magnetization measured in increasing (decreasing) applied field. 
The results, shown in the main panels of Figs.~\ref{figHysteresis}(a) and~\ref{figHysteresis}(b) for in-plane and out-of-plane field orientations, reveal ferromagnetic magnetization curves with a saturation magnetization of $\sim$ 320 emu/cm$^3$ at 4.5 K, corresponding to 6.7 $\mu_B$/Eu, slightly less than the expected full moment of 7 $\mu_B$/Eu.
The comparison of magnetization curves measured along the three crystal axes [Fig.~\ref{figHysteresis}(c)] reveals a clear anisotropy in the approach to saturation with the saturation fields of $H$ // (110) and (100) being substantially smaller than for $H$ // (001) while there is no discernable in-plane anisotropy.
However, since the sample is plate-like such in-plane versus out-of-plane anisotropy may arise simply from demagnetization effects.
The dashed lines in Fig.~\ref{figHysteresis} indicate the corresponding demagnetization fields obtained by approximating the sample as an ellipsoid, demonstrating that the intrinsic saturation fields are indeed anisotropic with $H_{sat}^{ab} \sim$ 2.1 kG and $H_{sat}^c \sim$ 4.2 kG, consistent with easy-plane magnetic anisotropy.
Eu$^{2+}$ has a spin-only magnetic moment, and therefore, crystal electric field effects are not important in determining the single-ion magnetic anisotropy.
In the case of Eu-122 it has been suggested \cite{XiaoBrueckel2010_PhysRevB.81.220406} that dipolar interactions give rise to the easy-plane magnetic anisotropy.

In a model of a type-A antiferromagnet, the in-plane magnetization curves for which demagnetization effects are negligible allow for an estimate of the antiferromagnetic interlayer exchange constant $J'$.  Neglecting a weak in-plane anisotropy, the magnetization curve for this orientation is given by $M/M_s = H/H_{af}$ \cite{BuschowDeBoer2003}, where $g\mu_B H_{af} = 2z'\lvert J'\rvert S$ defines the antiferromagnetic exchange field $H_{af}$, and $z' = 2$ is the number of nearest neighbors along the $c$-axis yielding $J' \sim -0.04$ K.
This value may be largely overestimated as $c >> a$ (see Fig.~\ref{figStructure}) and the distance to the next nearest neighbors along the $c$-axis is only 4\% larger than the nearest neighbor distance and therefore an estimate with $z' \sim 10~(J' \sim -0.01$ K) would be more realistic.
In comparison, the ferromagnetic in-plane exchange constant $J$, as estimated from the paramagnetic Curie-Weiss temperature $\Theta_C = 2[zJ+z'J']S(S+1)/3k_B$, is $\sim$ 0.6 K, underlining the quasi-2D nature of magnetism.
Here, $z = 4$ is the number of in-plane nearest neighbors.
We believe that these order-of-magnitude estimates of the anisotropic exchange interactions remain valid even if the magnetic structure is more complicated than type-A, such as helical, for instance.
As RbEuFe$_4$As$_4$ is metallic and the Eu-4f moments are well localized within the Eu-ion situated $\sim$ 2 eV below the Fermi energy \cite{Jeevan2008_PhysRevB.78.052502} the indirect RKKY exchange interaction has been proposed as the mechanism of magnetic coupling \cite{LiuCao2016_PRB-polycrytalline-RbEu1144}.
While strong in-plane exchange interactions could also arise from superexchange for instance through the As-site, exchange in the $c$-direction and the onset of three-dimensional magnetic order will inevitably involve the predominantly Fe-3d states on two intervening superconducting FeAs-layers.
Orbital-selective magnetic and superconducting interactions may facilitate this coupling where superconductivity involves mainly $d_{xz}$ and $d_{yz}$ states \cite{Sprau2017_Science,Raghu2008_PhysRevB.77.220503} while the $d_{3z^2-r^2}$ orbital may transmit magnetic coupling along the $c$-axis \cite{RenXu2009_PhysRevB.79.094426}.
However, a recent study on polycrystalline Ni-for-Fe subtituted RbEu(Fe$_{1-x}$Ni$_x$)$_4$As$_4$ has shown that the magnetic ordering temperature is essentially independent of doping even as superconductivity is suppressed and a SDW on the Fe-sites re-emerges \cite{LiuCao2017-PhysRevB.96.224510}.
These results suggest that the RKKY interaction may not be the dominant interaction, and that the microscopic mechanisms underlying the simultaneous presence of sizable magnetic exchange and superconducting pairing interactions are not fully understood yet.

\begin{center}
\textbf{C. Resistivity}
\end{center}

\begin{figure}
	\includegraphics[width=1\columnwidth]{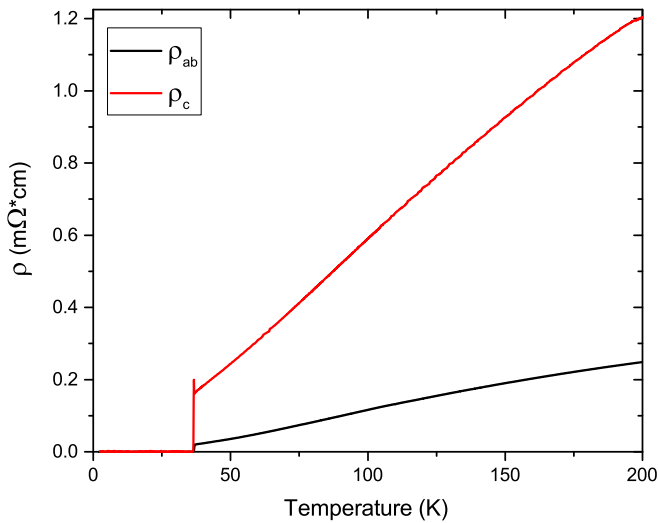}
	\caption{
		Temperature dependence of the in-plane and out-of-plane resistivities of RbEuFe$_4$As$_4$.
		The anisotropy changes from $\sim$4 at 200 K to $\sim$8 near $T_c$.
	}
	\label{figResistivity}
\end{figure}

Figure~\ref{figResistivity} shows the temperature dependence of the in-plane ($\rho_{ab}$) and $c$-axis ($\rho_c$) electrical resistivities of RbEuFe$_4$As$_4$ measured with $i$ = 1 mA.
The residual resistivity $\rho_{ab}(0)$ is estimated at approximately 15 $\mu\Omega$ cm, indicating high-quality material.
$\rho_{ab}$ and $\rho_c$ are metallic, a feature seen in other 1144-type and 122-type superconductors \cite{MeierCanfield2016_PhysRevB.94.064501,TanatarProzorov2009_PhysRevB.79.134528}.
The resistivity anisotropy increases from $\sim$ 4 near 200 K to about 8 at $T_{c}$, similar to the behavior of non-magnetic CaKFe$_4$As$_4$ \cite{MeierCanfield2016_PhysRevB.94.064501}.
Such temperature-dependent anisotropy could arise in a multi-band system in which carriers in the different bands have different mobilities with different temperature dependences.
In all samples studied, there is a sharp drop in the resistivity at the superconducting transition temperature of $T_{c} \sim$ 36.5 K to 36.8 K, with a transition width of 0.5 K or less.
The sharp feature at the top of the $c$-axis resistive transition arises from non-ideal contact geometry and the redistribution of the current flow at the superconducting transition \cite{Vaglio1993_PhysRevB.47.15302}.
Below $T_{c}$ down to 1.6 K, we do not observe a re-entrant resistive state associated with the onset of magnetic order of the Eu-sublattice, unlike observed in other Eu-containing iron arsenides such as Eu(Fe$_{1-x}$Ir$_x$)$_2$As$_2$ and EuFe$_2$As$_2$ (under pressure) \cite{KuritaTerashima2011_PhysRevB.83.214513,Paramanik2013_JoPCM-Eu(FeIr)2As2,Jiao2011_EPL} or in several borocarbide superconductors \cite{MullerNarozhnyi_RepProgPhys2001,Eisaki1994_PhysRevB.50.647}.
Our finding is consistent with very weak coupling of Eu-magnetism and superconductivity in RbEuFe$_4$As$_4$.

\begin{figure*} 
	\includegraphics[width=2\columnwidth]{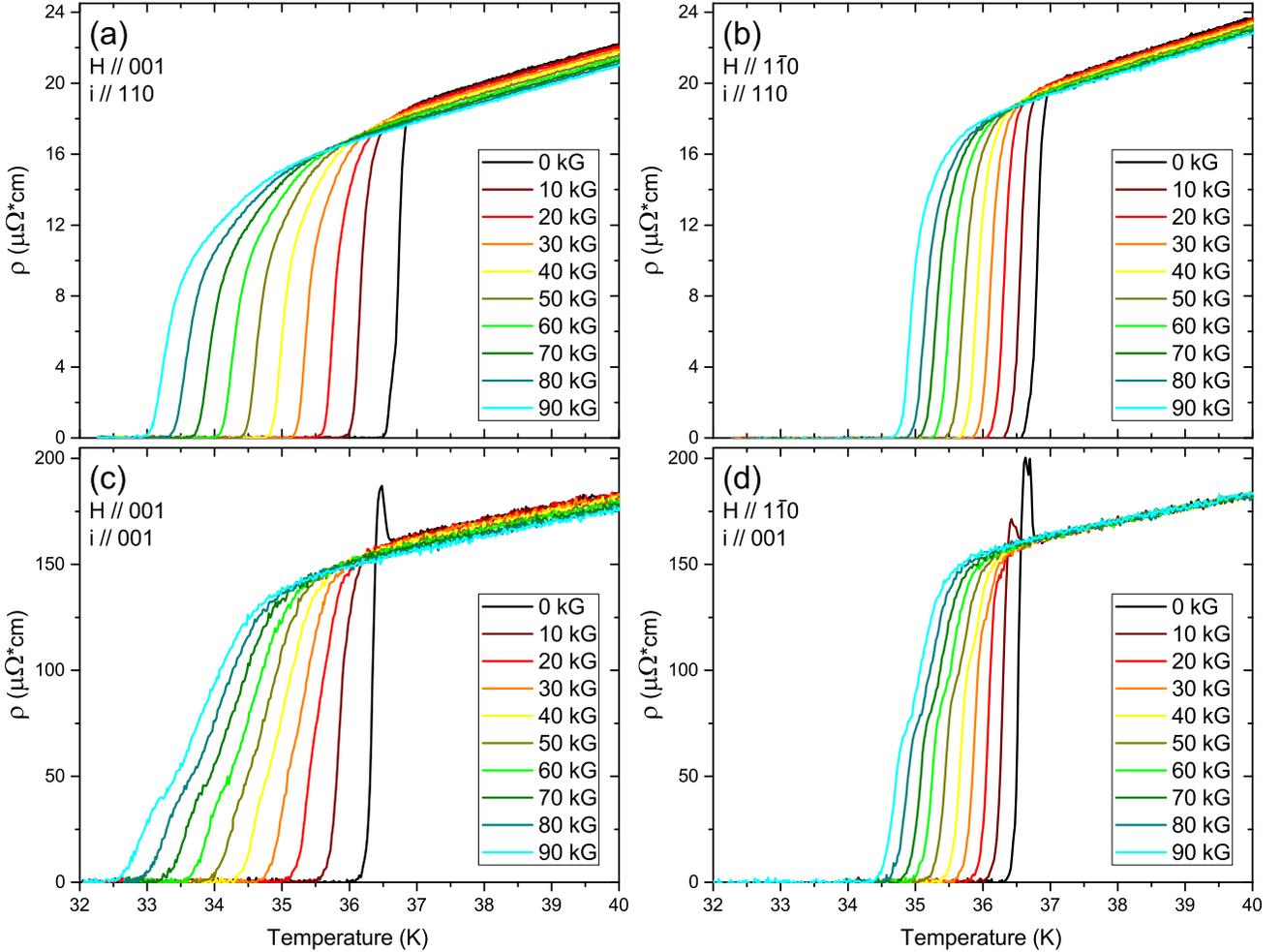}  
	\caption{
		Temperature and field dependence of the resistivity for various field and current configurations.
		(a) and (b) current in-plane and field applied along the $c$-axis and parallel to the $ab$-planes, respectively.
		(c) and (d) $c$-axis current and field applied along the $c$-axis and the $ab$-planes.   
	}
	\label{figRTH}
\end{figure*}

To study the superconducting anisotropy, resistivity measurements with applied magnetic field parallel to the (1$\bar{1}$0) ($ab$-plane) or the (001) ($c$-axis) directions were performed (Fig.~\ref{figRTH}).
Fig.~\ref{figRTH}(a) and~\ref{figRTH}(b) show resistivity measurements up to 9 T with an $ab$-planar current of 1 mA, with the field parallel to $c$ and parallel to $ab$, respectively, on the same single crystal of RbEuFe$_4$As$_4$.
In both cases, the field was perpendicular to the current.
A modest anisotropy, the rather large slopes of $H_{c2}(T)$, and negative normal-state magnetoresistance are all immediately noticeable.
Fig.~\ref{figRTH}(c) and~\ref{figRTH}(d) show resistivity measurements with 1 mA parallel to the $c$-axis on a separate crystal, with the field parallel to $c$ and $ab$, respectively.
The results are qualitatively the same as for $ab$ planar current.

\begin{figure} 
	\includegraphics[width=1\columnwidth]{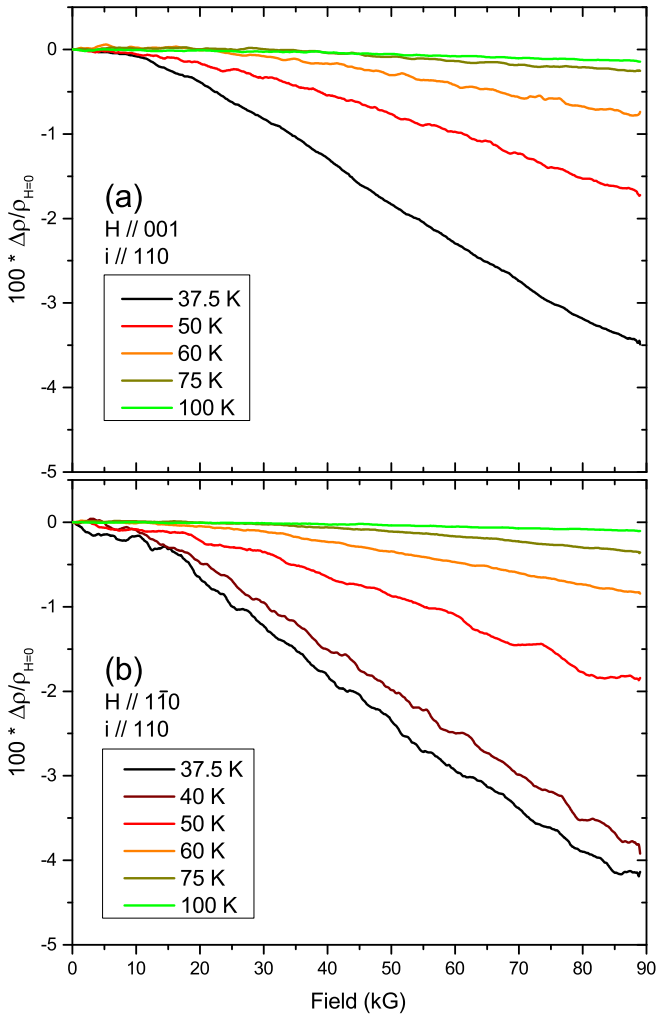}  
	\caption{
		Transverse magnetoresistance $\Delta\rho/\rho(H=0)$ at multiple fixed temperatures with $i$ // (110) for (a) $H$ // (001) and (b) $H$ // (1$\bar{1}$0) in a single crystal of RbEuFe$_4$As$_4$.   
	}
	\label{figRH}
\end{figure}

The resistivity data corresponding to the normal state in Fig.~\ref{figRTH} reveal a negative magnetoresistance (MR).
Fig.~\ref{figRH} shows measurements of the isothermal transverse MR, $\Delta\rho/\rho(H=0) = (\rho(H)-\rho(H=0))/\rho(H=0)$, for $H$ // (001) and $H$ // (1$\bar{1}$0) at various temperatures with current along (110).
For $H$ // (1$\bar{1}$0), the field and current were perpendicular.
The MR was obtained by slowly sweeping the applied magnetic field from -9 to 9 T and by evaluating the symmetric part of the signal in order to eliminate spurious contributions from the Hall effect in non-ideal contact geometries.
Measurements at currents of 1 mA and 0.1 mA yielded the same results.
We observe a clearly discernable negative transverse MR at temperatures above $T_{c}$, that is, in the paramagnetic state of the Eu-ions.
With increasing temperature the MR decreases rapidly.
A negative MR has been observed previously in EuFe$_2$As$_2$ \cite{Jiang2009_NJoP,Terashima2010_JPSJ-EuFe2As2} in the magnetically ordered and paramagnetic states of the Eu-sublattice, and has been attributed to the suppression of electron scattering by Eu-spin fluctuations.
An analysis based on the Yamada-Takada model \cite{YamadaTakada1973_JPSJ} yielded a quantitative description of the effect \cite{Terashima2010_JPSJ-EuFe2As2}.
The observation of a large magnetic contribution to the specific heat in high fields and at high temperatures [23, 24] reveals sizable spin-fluctuations at temperatures well above $T_m$ and that a similar mechanism of negative MR may be active in RbEuFe$_4$As$_4$.
The data at 37.5 K suggest a change in curvature of the MR at high fields indicating the superposition of two effects, the negative MR at low fields due to suppression of spin-scattering and the conventional positive MR due to the cyclotron motion of the carriers that grows as $(\mu H)^2$ where $\mu$ is the carrier mobility.
As in the case of EuFe$_2$As$_2$, the negative MR is a small effect, of the order of a few percent.

\begin{center}
\textbf{D. Superconducting phase diagram}
\end{center}

\begin{figure}
	\includegraphics[width=1\columnwidth]{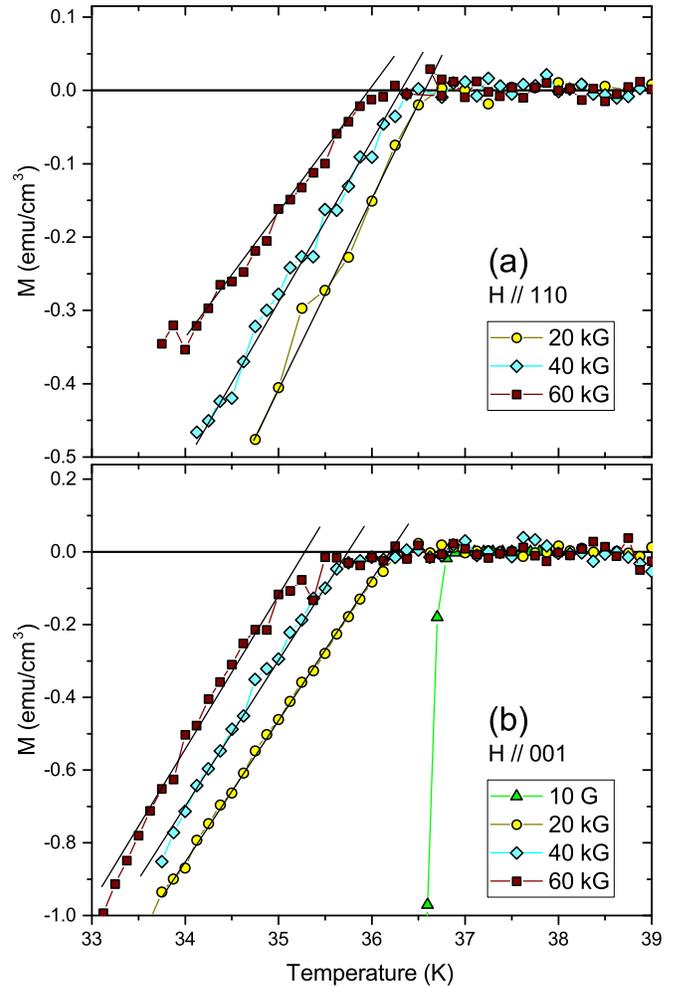}
	\caption{
		Temperature dependence of the magnetization in various fields applied along the $ab$-planes (a) and the $c$-axis (b).
		The lines indicate the construction of $T_{c}(H)$.
	}
	\label{figMagT}
\end{figure}

\begin{figure}
	\includegraphics[width=1\columnwidth]{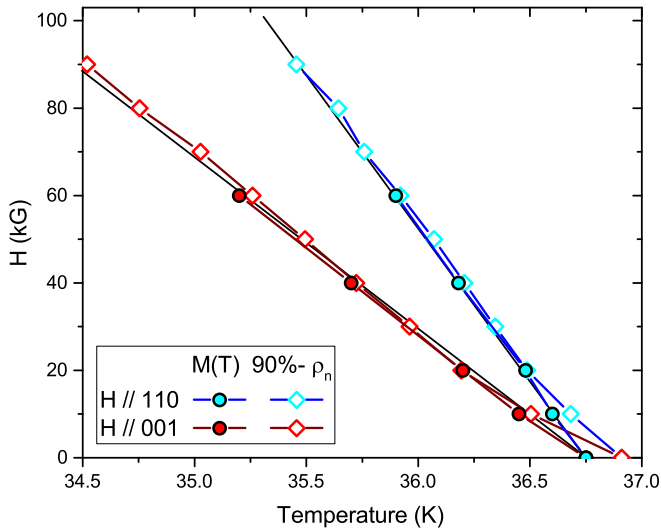}
	\caption{
		The upper critical field of RbEuFe$_4$As$_4$ as determined from magnetization (closed circles) and magnetotransport (open diamonds) measurements.  
	}
	\label{figPhase}
\end{figure}

We determine the superconducting phase diagram of RbEuFe$_4$As$_4$ from the resistive transitions shown in Fig.~\ref{figRTH} and from magnetization measurements in fields up to 60 kG, see Fig.~\ref{figMagT}. 
Here, a quadratic polynomial in $1/T$ has been fitted between 37 K and 40 K and subtracted from the magnetization data such as those shown in Fig.~\ref{figXT}(b) to reveal the superconducting signature.
The anisotropic shift of the superconducting transition in applied fields is clearly seen.
We observe that the 90\%-$\rho_n$ criterion and the magnetic determination yield consistent measures of $T_c(H)$.
The resulting phase boundaries are shown in Fig.~\ref{figPhase}.
We find enormous upper critical fields $H_{c2}$ and remarkably low superconducting anisotropies $\Gamma$, in line with the behavior generally seen for Fe-based superconductors.
We obtain $dH_{c2}^{ab}/dT = -70 $ kG/K, $dH_{c2}^{c}/dT = -42 $ kG/K, $\Gamma = 1.7$ (not including the upward curvature near $T_{c}$, which is not apparent in the magnetization data).
The value for the anisotropy is lower than expected on the basis of the resistivity anisotropy and a single-band Drude model for which $\Gamma \sim \sqrt{\rho_c/\rho_{ab}}$ suggestive of multi-band effects and potential gap anisotropy. 

Results obtained on the non-magnetic sister compound CaKFe$_4$As$_4$ \cite{MeierCanfield2016_PhysRevB.94.064501,2018arXiv180206150T} suggest that RbEuFe$_4$As$_4$ is a multi-band superconductor; however, as neither the inter and intra band pairing constants nor the details of the Fermi surface are known, we present an approximate discussion of the upper critical field using a single-band formalism.
Using the GL relationship $H_{c2}(0) = -(dH_{c2}/dT)\rvert_{T_{c}}*T_{c}$, we estimate zero-temperature values of $H_{c2}^{ab}(0) \sim$ 2500 kG and $H_{c2}^c(0) \sim$ 1600 kG, very large but comparable to other Fe-based superconductors \cite{Yuan2009_Nature,Gurevich2011}.
These estimates exceed the BCS paramagnetic limit $H_p$(kG) $= (1+\lambda)*18.4 T_{c}$(K) where $\lambda$ is the electron-boson coupling constant \cite{SchossmanCarbotte1989_PhysRevB.39.4210} even when including strong-coupling effects, indicating that at low temperatures deviations from the GL extrapolation will occur.
Nevertheless, the in-plane and out-of-plane GL coherence lengths $\xi_{ab}$ and $\xi_c$ may be estimated using the single-band Ginzburg-Landau relations $H_{c2}^c(0) = \Phi_0/2\pi\xi_{ab}^2(0)$ and $H_{c2}^{ab}(0) = \Phi_0/[2\pi\xi_{ab}(0)\xi_c(0)]$ yielding $\xi_c(0)$ = 0.92 nm and $\xi_{ab}$ = 1.4 nm.
The estimate for $\xi_c(0)$ is slightly smaller than the $c$-axis lattice constant making the low value of the anisotropy all the more surprising.

With the help of the Rutgers relation a connection between the jump in the specific heat and the superconducting phase boundaries can be established: $\Delta C/T_{c} = (dH_{c2}^i/dT\rvert_{T_{c}})^2/8\pi\kappa_i^2$.
With $\Delta C/T_{c}$ = 0.21 J/mol K$^2$ and the upper critical field slopes from Fig.~\ref{figPhase} we obtain the GL parameters $\kappa_c = \lambda_{ab}/\xi_{ab} \sim$ 67 and $\kappa_{ab} = \sqrt{\lambda_{ab}\lambda_c}/\sqrt{\xi_{ab}\xi_c} \sim$ 108, consistent with determinations based on the slopes of the $M(T)$ curves shown in Fig. \ref{figMagT}.
Thus, RbEuFe$_4$As$_4$ is in the extreme type-II limit, as is commonly observed for the Fe-based superconductors.

These materials parameters allow to estimate the Ginsburg number $G_i$, which describes the importance of superconducting thermal fluctuations, as $G_i = [8\pi^2 k_B \Gamma T_{c} \kappa_c^2 \xi_{ab}/\phi_0^2]^2/2 \approx 7 \times 10^{-5}$.
This relatively low value is of the same order of magnitude as seen in other 122 and 1144 Fe-based superconductors \cite{MeierCanfield2016_PhysRevB.94.064501,Chaparro2012_PhysRevB.85.184525}, and is consistent with the almost complete absence of fluctuation effects at the superconducting transition, see inset of Fig.~\ref{figCT}.
In contrast, $G_i$ is significantly larger in the 1111 compounds, $G_i \sim 10^{-3} - 10^{-2}$ \cite{Welp2008_PhysRevB.78.140510,Pribulova2009_PhysRevB.79.020508,Welp2011_PhysRevB.83.100513}, the principal difference being the much larger anisotropy of the 1111 materials.

\begin{center}
\textbf{IV. CONCLUSION} 
\end{center}

In summary, among the superconductors containing ordered sublattices of rare-earth magnetic moments RbEuFe$_4$As$_4$ attains a special place due to its high magnetic and superconducting transition temperatures.
Orbital-selective superconducting pairing and magnetic exchange may offer a frame for the coexistence of strong superconducting pairing and sizable magnetic interactions in this layered material even though the underlying microscopic mechanisms have not been clarified yet.
The high value of $T_{c}$, exceeding that of the non-magnetic sister compound CaKFe$_4$As$_4$ \cite{MeierCanfield2016_PhysRevB.94.064501}, doping studies \cite{LiuCao2017-PhysRevB.96.224510,Kawashima2018-LT28Proceedings} and the surprisingly low value of the superconducting anisotropy, $\Gamma \sim$ 1.7, indicate that both interaction channels are largely decoupled.
In contrast to superconductivity, Eu-magnetism is highly anisotropic quasi-two dimensional, reflecting the large separation between the Eu-layers.
We estimate in-plane and out-of-plane exchange constants of 0.6 K and less than 0.04 K, respectively.
This reduced dimensionality induces strong magnetic fluctuations, a sizable suppression of the magnetic ordering temperature below the paramagnetic Curie-Weiss temperature and a cusp-like specific heat anomaly devoid of any singular behavior.
These features distinguish RbEuFe$_4$As$_4$ from the parent compound EuFe$_2$As$_2$ in which the distance between Eu-layers is half, and magnetism is more three-dimensional-like.
Magnetization curves reveal a clear magnetic easy-plane anisotropy of RbEuFe$_4$As$_4$ with in-plane and out-of-plane saturation fields of 2 kG and 4 kG, respectively.

\begin{center}
\textbf{ACKNOWLEDGMENTS}
\end{center}
This work was supported by the U.S. Department of Energy, Office of Science, Basic Energy Sciences, Materials Sciences and Engineering Division.
This research used resources of the Advanced Photon Source, a U.S. Department of Energy (DOE) Office of Science User Facility operated for the DOE Office of Science by Argonne National Laboratory under Contract No. DE-AC02-06CH11357.
M.P.S. thanks ND Energy for supporting his research and professional development through the NDEnergy Postdoctoral Fellowship Program.
K.W. acknowledges support through an Early Postdoc Mobility Fellowship of the Swiss National Science Foundation.
Z.D. acknowledges support from the Swedish Research Council (VR) under Grant No. 2015-00585, co-funded by Marie Sk\l{}odowska-Curie Actions (Project No. INCA 600398).
A.R. acknowledges support from the Swedish Research Council (VR) under Grant No. 2016-04516.
The authors thank Roland Willa for helpful discussions.

\bibliography{RbEu1144_v1.27}

\end{document}